\documentclass[twocolumn,showpacs,prl,superscriptaddress,floatfix]{revtex4}

\usepackage[usenames]{color} 
\usepackage{ulem}
\usepackage{tabularx} 
\usepackage{epsfig}
\usepackage{amsmath} 
\usepackage{amssymb} 
\usepackage{graphicx}
\usepackage{goodfloat}

\begin{document}

\title{Evidence for universal scaling {\em in} the spin-glass phase}

\author{Thomas J\"org} \affiliation{LPTMS, Universit\'{e} Paris-Sud, CNRS UMR
  8626, 91405 Orsay Cedex, France}

\author{Helmut G.~Katzgraber} \affiliation {Theoretische Physik, ETH
  Z\"urich, CH-8093 Z\"urich, Switzerland}

\begin{abstract}

We perform Monte Carlo simulations of Ising spin-glass models
in three and four dimensions, as well as of Migdal-Kadanoff spin
glasses on a hierarchical lattice. Our results show strong evidence
for universal scaling in the spin-glass phase in all three models. Not
only does this allow for a clean way to compare results obtained from
different coupling distributions, it also suggests that a so far
elusive renormalization group approach within the spin-glass phase
may actually be feasible.

\end{abstract}


\normalem
\maketitle

The characterization of the spin-glass phase in finite space
dimensions remains as one of the prominent unresolved problems
in the physics of disordered systems. Although the spectrum
of theories that have been proposed to describe spin glasses is broad
\cite{parisi:79,mezard:84ea,bray:86,fisher:86,binder:86,huse:87,
newman:96,krzakala:00,palassini:00},
the main theoretical pictures are the simple scaling approach (droplet
picture) \cite{bray:86,fisher:86} and the replica symmetry breaking
(RSB) scenario inspired by mean-field theory \cite{parisi:79}.
Because frustration and disorder are inherent ingredients of
spin glasses, progress via field-theoretical calculations has been
difficult, at least below six dimensions \cite{dedominicis:06}. Thus
most of the progress in the field relies on numerical studies that
are also strongly limited mainly due to the numerical complexity of
spin glasses.  In fact, the difficulties are such that there is no
good numerical evidence of whether a renormalization group
(RG) approach may work in the spin-glass phase.

We address some of the necessary features that one should observe
in the spin-glass phase such that scaling theory and an RG approach
may be potentially successful. To check for the applicability of
a scaling theory within the spin-glass phase we assume {\em a priori}
that a scaling approach works and check {\em a posteriori} whether our
results are consistent with the scaling assumptions made. For this
purpose we study finite-size scaling functions where we compare the
behavior of several observables on a change of scale: We rescale
the system size and plot observables as a function of a
phenomenological coupling which in our case is the Binder cumulant
\cite{binder:81}.  We find that this procedure leads to
scaling functions consistent with a universal scaling behavior in
the spin-glass phase for all models studied: the Migdal-Kadanoff
(MK) spin glass on a (three-dimensional) hierarchical lattice
\cite{berker:79}, as well as the three- (3D)
and the four-dimensional (4D) Edwards-Anderson (EA) short-range Ising
spin glasses \cite{edwards:75}. Of particular interest is the scaling
function of the Binder cumulant itself since it provides a compact
way to look at the complete RG flow.

The MK spin glass serves as an example to illustrate the behavior of
the different finite-size scaling functions in a simple scaling theory.
This, in turn, follows from the possibility to exactly solve the model
(numerically) using an RG decimation transformation.

The physically interesting case of the 3D EA model is also the
most difficult as the lower critical dimension is close
\cite{boettcher:05d}, i.e., that the spin-glass phase in
this case is rather marginal. Hence we also study the model in 4D.
We observe similarities as well as clear differences between the
MK and EA models.  In particular, for all studied models we find
clear evidence for the emergence of a universal scaling behavior
in the spin-glass phase in the thermodynamic limit.  Although our
findings might be taken as a hint for the correctness of a simple scaling
approach for the EA model, it is fair to observe that for very low 
temperatures we find an effective stiffness exponent $\theta$ 
which is compatible with zero, as expected in the
RSB or TNT \cite{krzakala:00,palassini:00} scenarios.  Furthermore,
for the system sizes studied, the fractal dimension of the surface of
the low-energy excitations seems not to be equal to the space dimension
$d$, in agreement with the TNT or droplet scenarios.  We therefore lay
the foundation for a simple approach that should allow future studies
to check whether the disagreement with the traditional pictures is
due to scaling corrections, or whether new theoretical descriptions
are needed.

\paragraph*{Finite-size scaling approach.---}
\label{sec:scaling}

For any observable ${\mathcal{O}}(L,T)$ as a function of the
temperature $T$ and the system size $L$, and the finite-size
correlation length $\xi(L,T)$ finite-size scaling (FSS) theory
predicts \cite{privman:90} that
\begin{equation}
  {{\mathcal O}(s L,T)}/{{\mathcal O}(L,T)}=
  F_{{\mathcal O}}\!\left[ \xi(L,T)/L; s \right] +
   ({\rm corrections}), \notag 
  \label{eq:FSS_1}
\end{equation}
where $s$ is a scaling factor.  The corrections vanish in the
thermodynamic limit.  Because even the correlation length divided
by the system size diverges in the spin-glass phase, this definition
is inconvenient to examine the behavior of FSS functions within the
spin-glass phase. Hence we use an alternative phenomenological
coupling as the scaling variable which allows for a better
visualization of the scaling functions in the spin-glass phase. We
find that the Binder cumulant $g(L,T)$ [see Eq.~(\ref{eq:Binder}) below]
works best since it is bounded in the interval $[0,1]$, thus leading
to a compact picture of the scaling properties in the whole
spin-glass phase. In the following we study different FSS functions
\begin{equation}  
  {{\mathcal O}(s L,T)}/{{\mathcal O}(L,T)} = F_{{\mathcal O}} \!\left[
  g(L,T); s \right] + ({\rm corrections}) 
  \label{eq:FSS_2}
\end{equation}
defined as a function of $g(L,T)$. Next, we verify numerically whether
it is sensible to study such FSS functions in the spin-glass phase
and discuss our findings.

\paragraph*{Models and observables.---}
\label{sec:model}

We consider the spin-glass Hamiltonian ${\mathcal H} = - \sum_{i,
j} J_{ij} \sigma_i \sigma_j$, where the sum is over all nearest
neighbor pairs on the lattice and $\sigma_i \in \{\pm 1\}$ are Ising
spins. The couplings $J_{ij}$ are drawn from either a Gaussian
(G), bimodal (B), link-diluted bimodal (D), or irrational (I)
distribution which corresponds to a bimodal distribution where half
of the (randomly chosen) bonds are multiplied by an irrational
constant $c_{\rm I}=(1+\sqrt{5})/2$.  The Hamiltonian is studied both
on a hierarchical lattice with an effective space dimension of three
and on a simple hypercubic lattice in three and four space dimensions.
We use periodic boundary conditions for the EA model on the hypercubic
lattices and free boundary conditions for the hierarchical
lattice \cite{comment:bc}. The space dimension of the hierarchical 
lattice with $G$ generations is $d = 1 + \ln(b)/\ln(s)$, where $b$ 
is the number of parallel branches and $s$ is the number of bonds 
in series (we set $s = 2$ and $b = 4$ to obtain an effective space 
dimension of $3$). The size of the system is $L = s^G$.

The spin-glass order parameter is given by $q = (1/N)\sum_{i}
\sigma_i \tau_i$, where $\sigma$ and $\tau$ are two replicas of
the system with the same disorder. For the hierarchical lattice
we use \cite{gardner:84} $q = 1/(2 N_L)\sum_{\langle i j\rangle}
(\sigma_i \tau_i +  \sigma_j \tau_j)$ where the sum runs over all
links $N_L$ of the lattice.
The spin-glass susceptibility $\chi$ is defined via $\chi(L,T) = N
[\langle q^2\rangle]_{\rm av}$, where $\langle \, \cdot \, \rangle$
represents a thermal average and $[ \, \cdot \, ]_{\rm av}$ a disorder
average.  The Binder cumulant $g$ is defined as:
\begin{equation}
  \label{eq:Binder}
  g(L,T) = \frac{1}{2} \left( 3 - \frac{[\langle q^4\rangle]_{\rm
        av}} {[\langle q^2\rangle]_{\rm av}^2} \right)\!\!.
\end{equation}
Finally, we also study the link-overlap $q_l = (1/N_L) \sum_{\langle
i j\rangle} \sigma_i\sigma_j\tau_i\tau_j$, where the sum is over
all links. Within the TNT picture the fractal dimension $d_s$ of
large-scale excitations can be extracted from the variance of the link
overlap $\sigma^2_{ql}(L,T) = [\langle q_l^2\rangle - \langle q_l
\rangle^2]_{\rm av} \sim L^{-\mu_l}$, where $\mu_l = \theta + 2(d -
d_s)$ varies with temperature \cite{katzgraber:01ea}. The stiffness
exponent $\theta$ follows from the temperature derivative of the
finite-size scaling function of the Binder cumulant $F_{{\partial_{T}
g}}$ via the quotient method \cite{ballesteros:97ea}:
\begin{equation}
  s^{-\theta} = 1 + g^* \partial_g F_{g}(g,s) \big|_{g=g^*} 
  + {\rm corrections} ,
  \label{eq:theta}
\end{equation}
where $g^*$ is the value of the Binder cumulant at any given
strong-coupling fixed-point \cite{comment:fp} and $s$ is the
scale factor used in the definition of the scaling function. Note
that cumulants of the order parameter as well as the correlation
length $\xi/L$ are RG invariant quantities often referred to as
``phenomenological couplings.''  In contrast to traditional scaling
analyses of the spin-glass phase which use {\em bare} (unrenormalized)
couplings (e.g., temperature), here we use renormalized couplings
(e.g., Binder ratio) as scaling variables, thus presenting a cleaner
analysis.
\begin{table}[t]
\caption{
Parameters for the simulations of the 3D model with Gaussian
(3DG) and irrational (3DI) disorder, as well as the 4D model with
Gaussian (4DG) and bond-diluted (4DD) disorder. $L$ is the system size,
$N_{\rm sa}$ is the number of disorder realizations, $N_{\rm sw}$
is the number of equilibration and measurement sweeps, $T_{\rm min}$ is 
the lowest temperature and $N_{r}$ the number temperatures used in the 
exchange Monte Carlo method.
\label{tab:simparams}}
{\footnotesize
\begin{tabular*}{\columnwidth}{@{\extracolsep{\fill}} l@{\hspace{-2.25em}}r r r c c }
\hline
\hline
Model &     $L$  &  $N_{\rm sa}$  & $N_{\rm sw}$ & $T_{\rm min}$ & $N_{r}$\\
\hline
3DG & $4$  & $109212$ & $1048576$  & $0.20$ & $22$ \\ 
 & $5$  & $100303$ & $1048576$  & $0.20$ & $22$ \\
 & $6$  & $101643$ & $1048576$  & $0.20$ & $22$ \\
 & $8$  & $40430$  & $8388608$  & $0.20$ & $22$ \\
 & $10$ & $10687$  & $33554432$ & $0.20$ & $22$ \\
 & $12$ & $5134$   & $33554432$ & $0.42$ & $18$ \\
 & $16$ & $5003$   & $8388608$  & $0.50$ & $17$ \\
\hline
3DI & $4$  & $160000$ & $256000$   & $0.30$ & $17$ \\
 & $5$  & $160000$ & $256000$   & $0.30$ & $17$ \\
 & $6$  & $160000$ & $512000$   & $0.30$ & $17$ \\ 
 & $8$  & $160000$ & $1024000$  & $0.30$ & $17$ \\
 & $10$ & $23712$  & $4096000$  & $0.30$ & $26$ \\
 & $12$ & $12768$  & $4096000$  & $0.70$ & $22$ \\
\hline
4DG & $3$  &  $20000$ &  $131072$ & $1.40$ & $29$ \\
 &      & $100000$ &   $16384$ & $0.39$ & $20$ \\
 & $4$  &  $20000$ &  $131072$ & $1.40$ & $29$ \\
 & $5$  &  $20000$ &  $131072$ & $1.40$ & $29$ \\
 & $6$  &  $20000$ &  $131072$ & $1.40$ & $29$ \\
 &      &  $10025$ & $4194304$ & $0.39$ & $20$ \\ 
 & $8$  &   $3500$ &  $524288$ & $1.40$ & $29$ \\
 & $10$ &   $2000$ &  $524288$ & $1.40$ & $29$ \\
\hline
4DD & $3$  &  $11392$ & $100000$ & $0.50$ & $11$ \\
 &      & $102400$ & $20000$  & $0.95$ & $11$ \\
 & $4$  & $107680$ & $40000$  & $0.95$ & $11$ \\
 & $5$  & $101699$ & $40000$  & $0.95$ & $11$ \\
 & $6$  &   $3072$ & $200000$ & $0.50$ & $11$ \\
 &      & $101664$ & $40000$  & $0.95$ & $11$ \\
 & $8$  &  $41408$ & $100000$ & $0.95$ & $21$ \\
 & $10$ &  $24160$ & $100000$ & $0.95$ & $21$ \\
\hline
\hline
\end{tabular*}
}\vspace{-0.4cm}
\end{table}

\paragraph*{Computational details.---}
\label{sec:numerical}

For the hierarchical lattice we use a similar procedure to
Ref.~\cite{sasaki:03}.  An alternative way to calculate the
link- and spin-overlaps on the hierarchical lattice is given in
Ref.~\cite{moore:98ea}. For $L \le 16$ ($G \le 4$) we use $N_{\rm
sa} = 5\cdot10^5$ samples, for $L = 32$ ($G = 5$) at least $10^5$ 
samples, for $L = 64$ ($G = 6$) at least $9\cdot10^4$ samples and for
$L = 128$ ($G = 7$) at least $1.5\cdot10^4$ samples.  All data
are averaged over $10^3$ independent configurations.

For the regular lattices the simulations are performed using exchange 
Monte Carlo \cite{hukushima:96, comment:cluster}. For systems with Gaussian 
disorder we use the equilibration test of Ref.~\cite{katzgraber:01ea}, 
whereas for the link-diluted case we perform a logarithmic binning of 
the data. Once the last three bins agree within error bars the system 
is equilibrated. For the 3D model with irrational [Gaussian] couplings $T_c = 1.47(3)$
[$T_c = 0.951(9)$]. For the 4D link-diluted model with 35\% dilution \cite{joerg:08c}
[Gaussian disorder] $T_c = 1.0385(25)$ [$T_c = 1.805(10)$]. For details
see Table \ref{tab:simparams}.
\paragraph*{Finite-size scaling functions.---}
\label{sec:scaling_functions}

\begin{figure}[b]

\vspace*{-0.60cm}

\includegraphics[width=0.83\columnwidth]{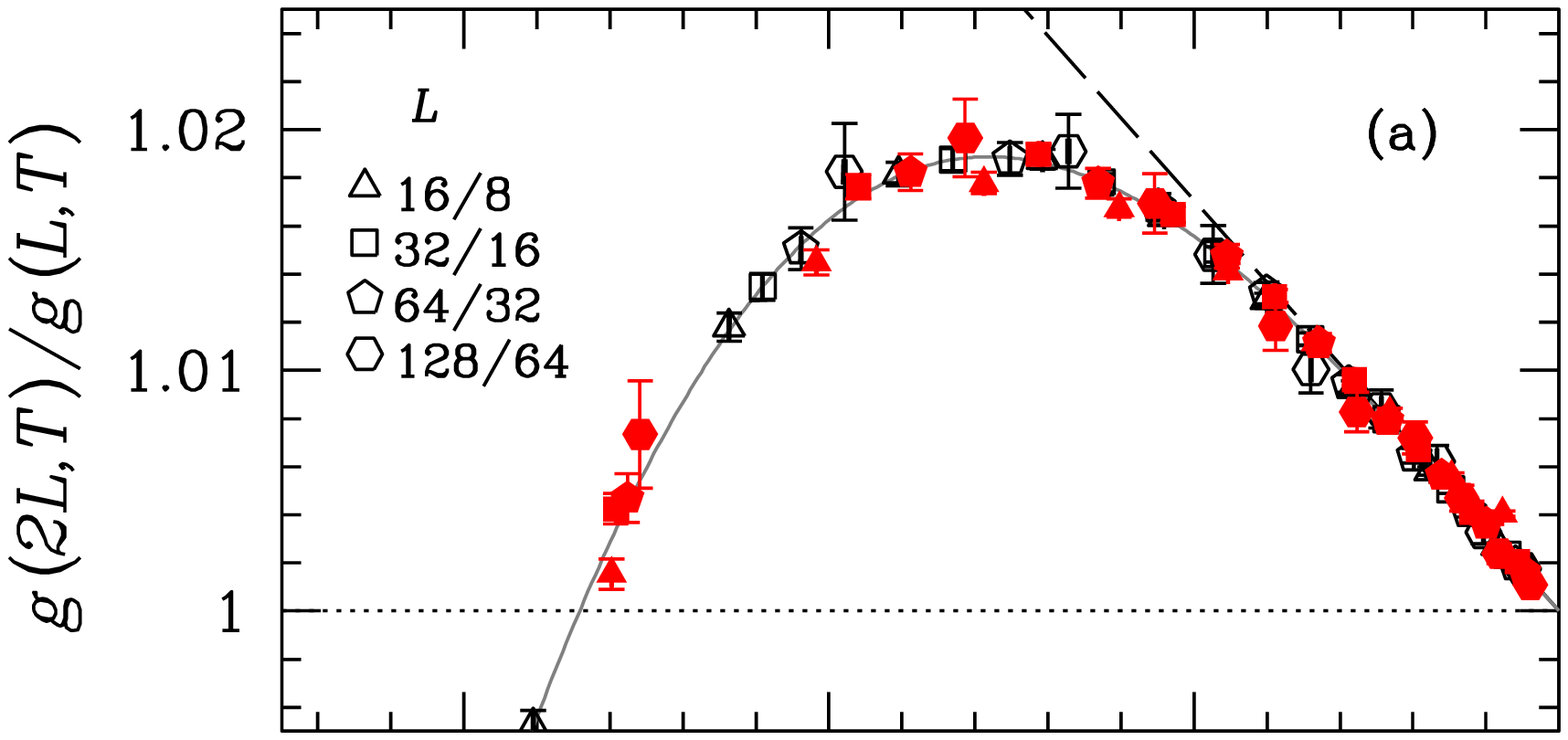}

\vspace*{-0.25cm}

\includegraphics[width=0.83\columnwidth]{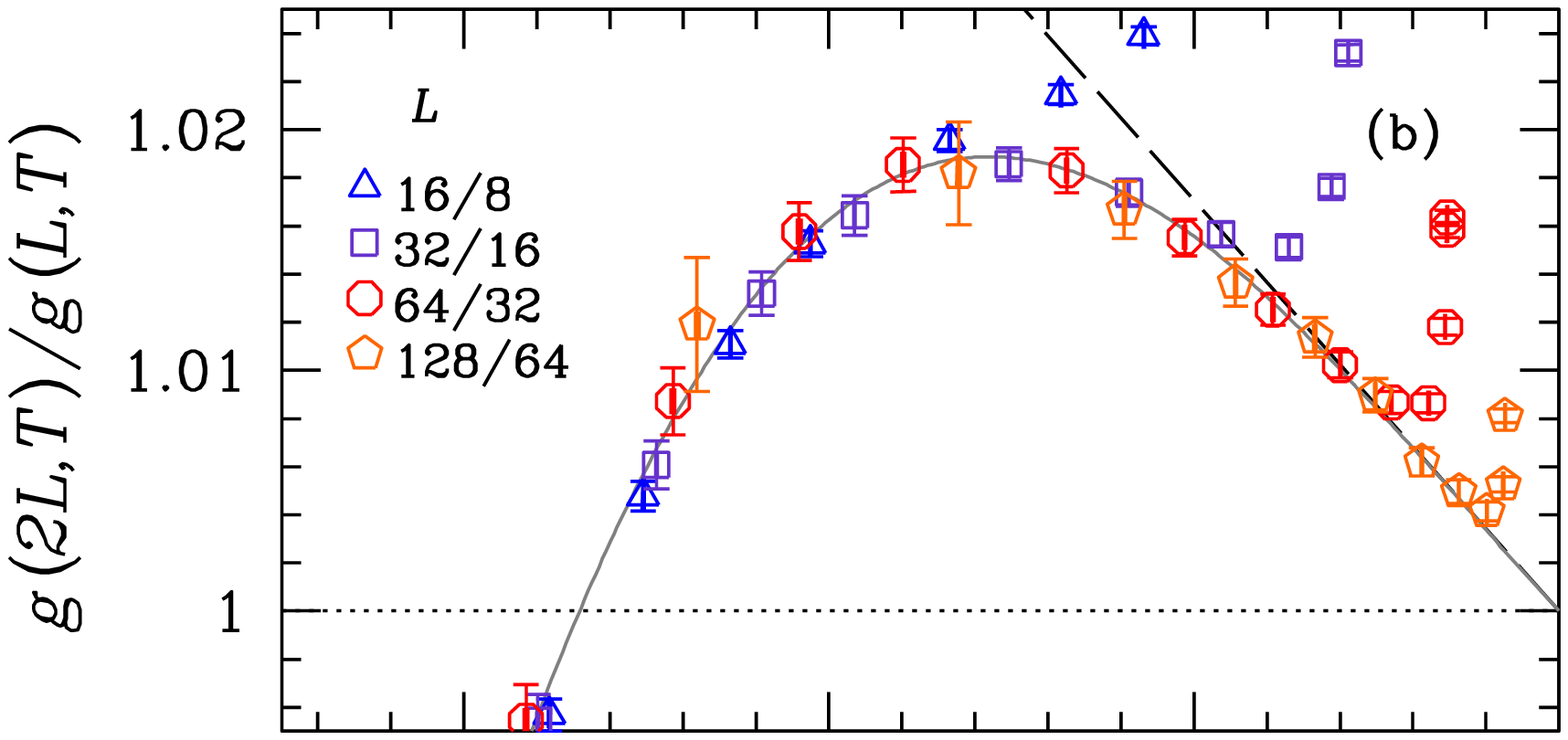}

\vspace*{-0.25cm}

\includegraphics[width=0.83\columnwidth]{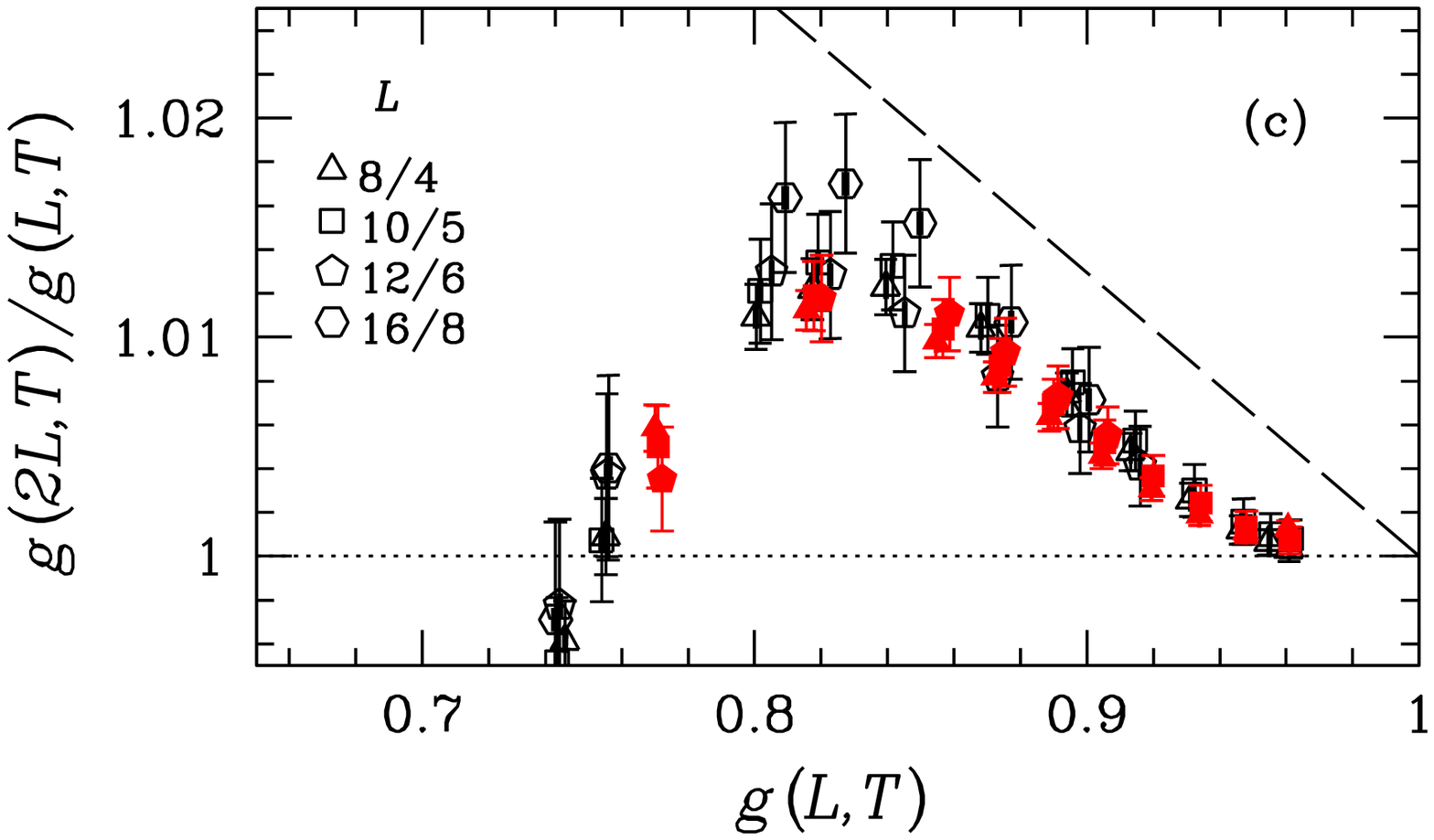}

\vspace*{-0.1cm}

\includegraphics[width=0.83\columnwidth]{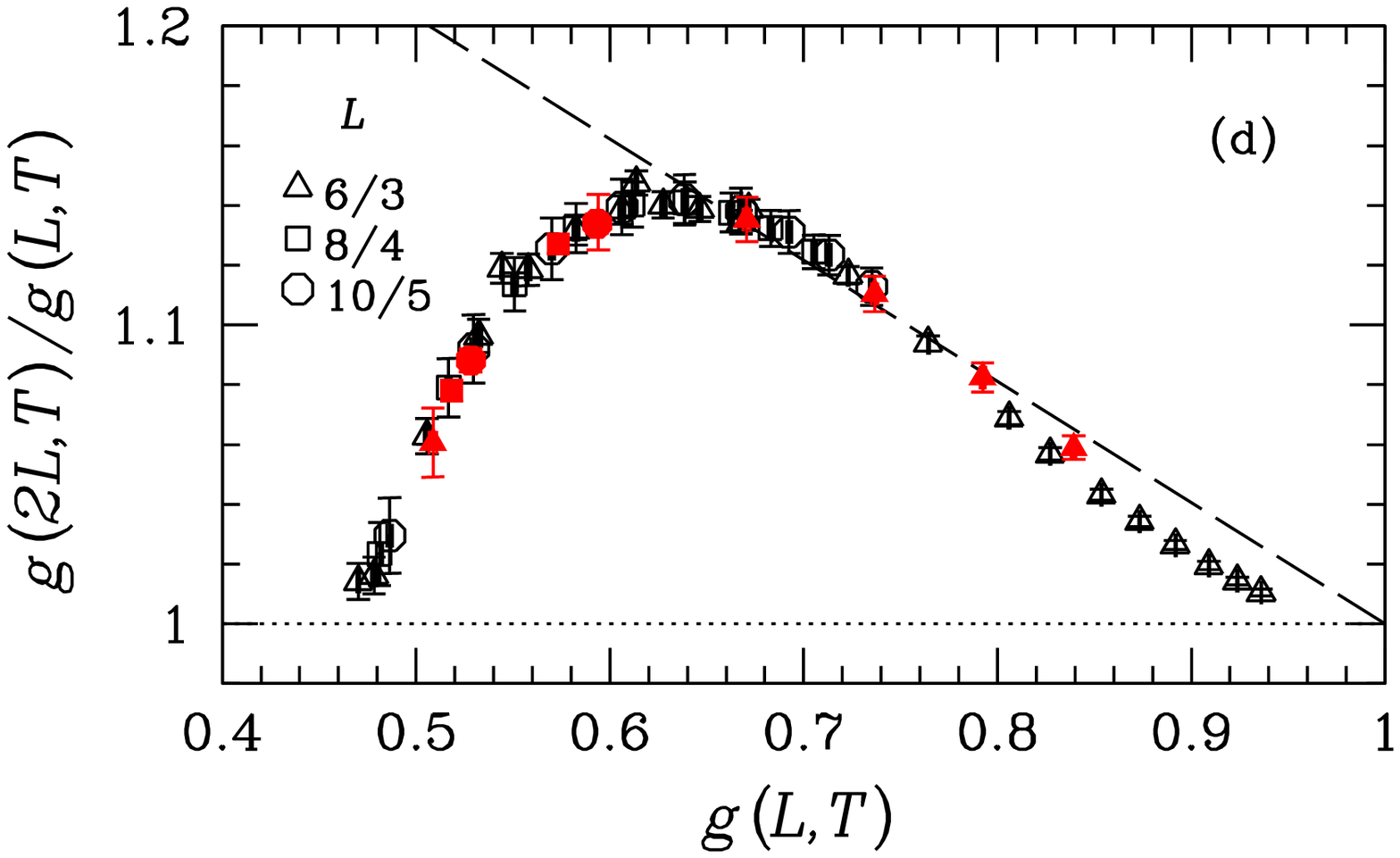}

\caption{(Color online) 
Comparison of the finite-size scaling function of the Binder
cumulant for different models. (a) Gaussian (open/black symbols) and
irrational (full/red symbols) disorder for the hierarchical lattice.
(b) Hierarchical lattice with bimodal disorder. The data converge
slowly to the Gaussian limiting case (solid curve). (c) 3D EA model
with Gaussian (open/black symbols) and irrational (full/red symbols)
disorder. (d) 4D EA model with both link-diluted (full/red symbols)
and Gaussian (open/black symbols) disorder. The dashed lines in all
panels represent a droplet scaling behavior with $\theta = 0.27$ for 
the hierarchical, $\theta = 0.2$ for the 3D and $\theta = 0.75$ for 
the 4D models, respectively.}
\label{fig:Binder_scaling}
\end{figure}
We first address the scaling function of the Binder cumulant $F_g$
for all models studied and then discuss further scaling functions.
In Fig.~\ref{fig:Binder_scaling} we show our results for the FSS
function $F_g$ for different models. $F_g$ displays in a compact way
the RG flow of the Binder cumulant. Panels (a) and (b) show results
for the hierarchical model where the scaling approach is known to work.
Panel (a) shows a comparison between Gaussian and irrational coupling
distributions, whereas panel (b) shows how the FSS scaling function
for the bimodal coupling distribution---amidst strong finite size
effects---converges (slowly) towards the FSS function of the Gaussian
(and irrational) cases.  The convergence is slowest close to $g =
1$ ($T = 0$) because for low $T$ entropic effects become relevant \cite{joerg:08a}.
These results are a clear indication for universal FSS {\em in} the
spin-glass phase for the MK model. Panel (c) shows a comparison of
the Gaussian and the irrational coupling distribution in the 3D EA
model. Finally, panel (d) shows a comparison between the Gaussian and
the link-diluted bimodal EA model in 4D. The results for the 3D/4D
EA models are consistent with the scaling hypothesis and indicate a
universal scaling behavior. The broken lines indicates how the FSS
function should depart from $g = 1$ assuming that simple droplet
scaling is correct, i.e., that the slope is given by the exponent
$\theta$ through Eq.~(\ref{eq:theta}). While this works perfectly for
the MK model, there is a clear difference in the case of the EA model.

\begin{figure}[b]

\vspace*{-0.6cm}

\includegraphics[width=0.83\columnwidth]{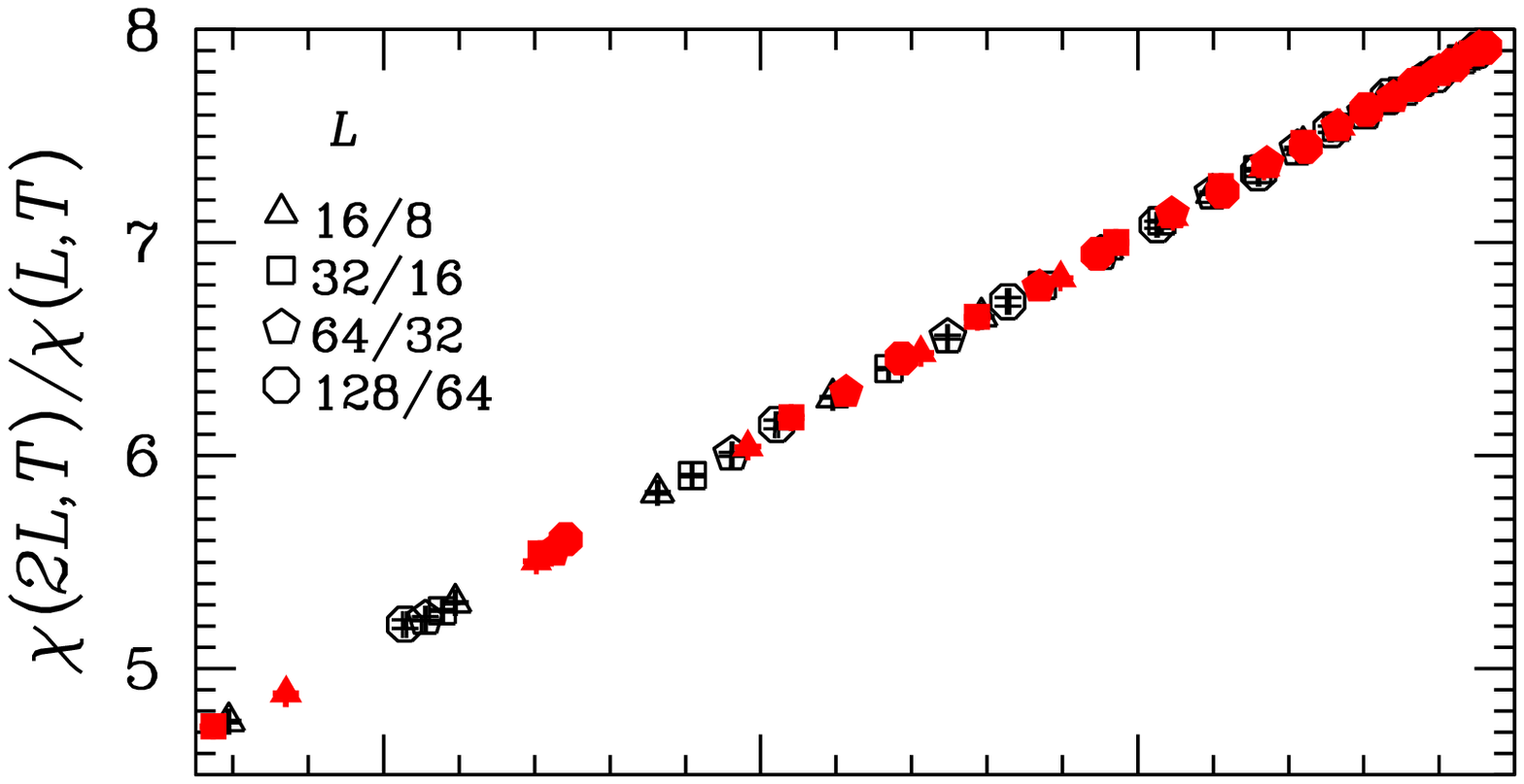}

\vspace*{-0.25cm}

\includegraphics[width=0.83\columnwidth]{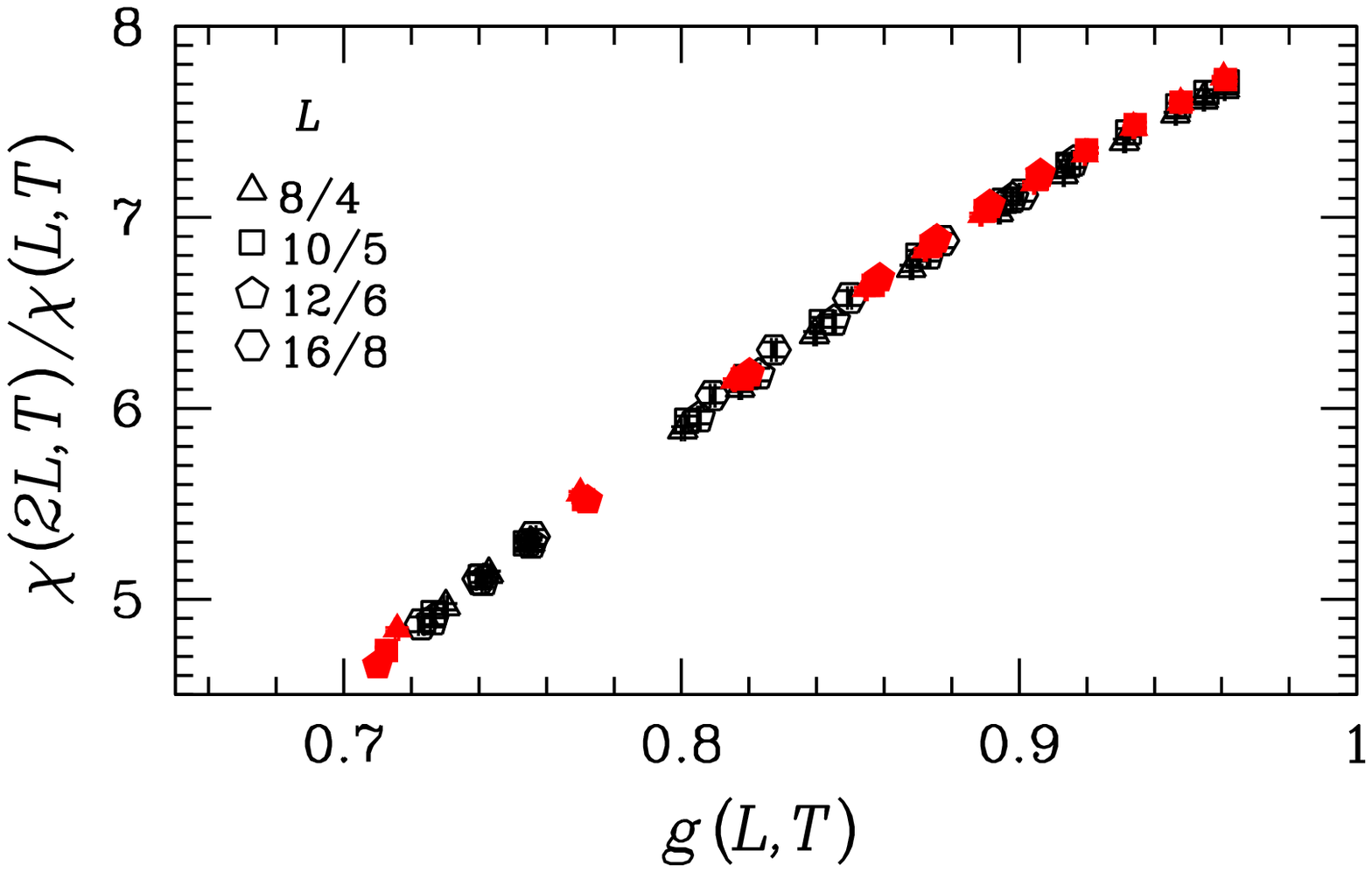}

\caption{(Color online) 
Comparison of the finite-size scaling function of the susceptibility
$\chi$ for the hierarchical lattice (top) with the 3D EA model
(bottom). In both models the results for the Gaussian (open/black
symbols) and the irrational (full/red symbols) coupling distribution
are shown.}
\label{fig:Chi_scaling}
\end{figure}

In order to show that the scaling collapse in
Fig.~\ref{fig:Binder_scaling} is not coincidental we show in
Fig.~\ref{fig:Chi_scaling} the FSS function of the spin-glass
susceptibility. The top panel shows a comparison of Gaussian and
irrational disorder for the MK model, whereas the bottom panel shows a
comparison for the 3D EA model. Again, the data show strong evidence
of a universal scaling behavior. The data for the MK model seem to
fall onto a straight line which underlines the very simple scaling
behavior in this model.

\begin{figure}[b]

\vspace*{-0.5cm}

\includegraphics[width=0.83\columnwidth]{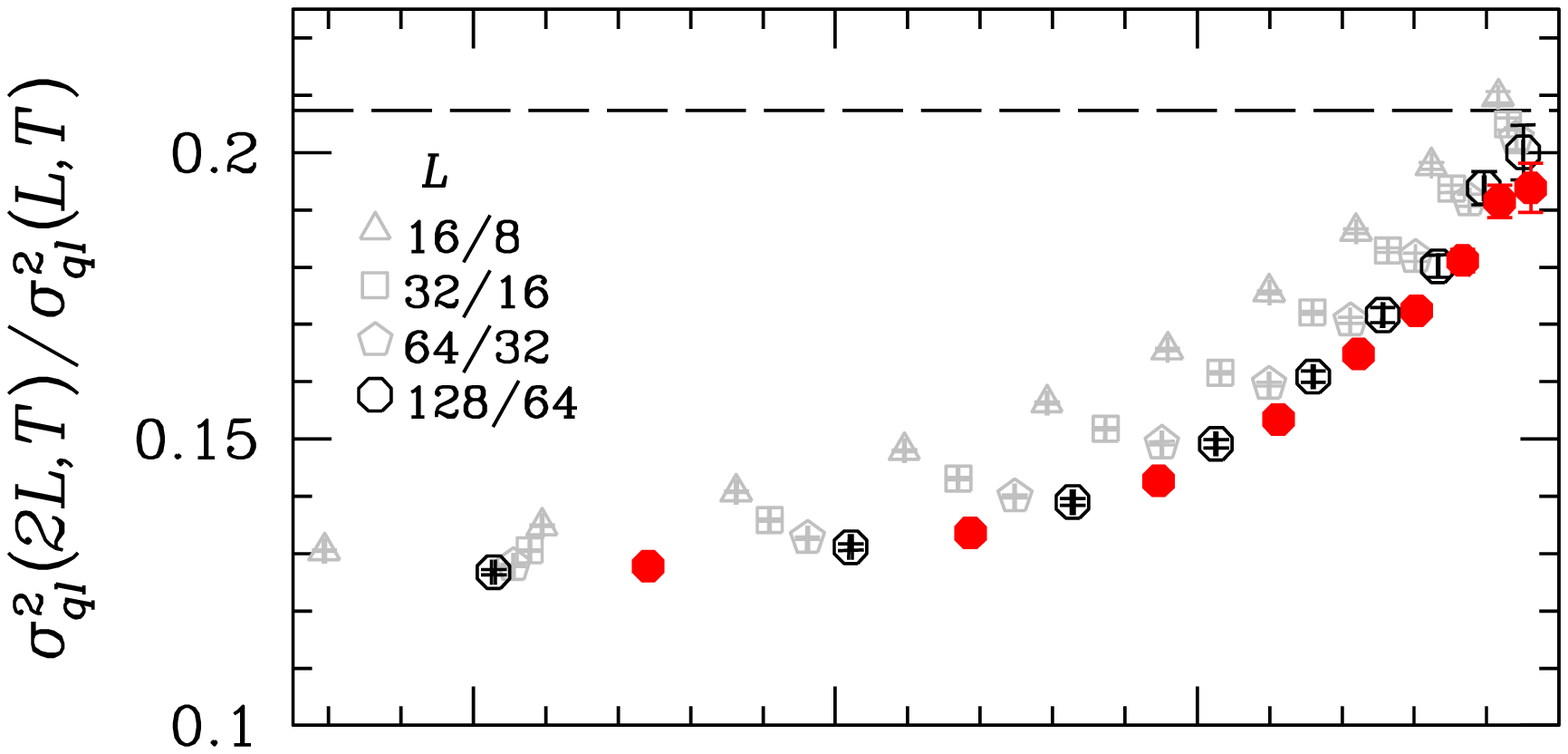}

\vspace*{-0.25cm}

\includegraphics[width=0.83\columnwidth]{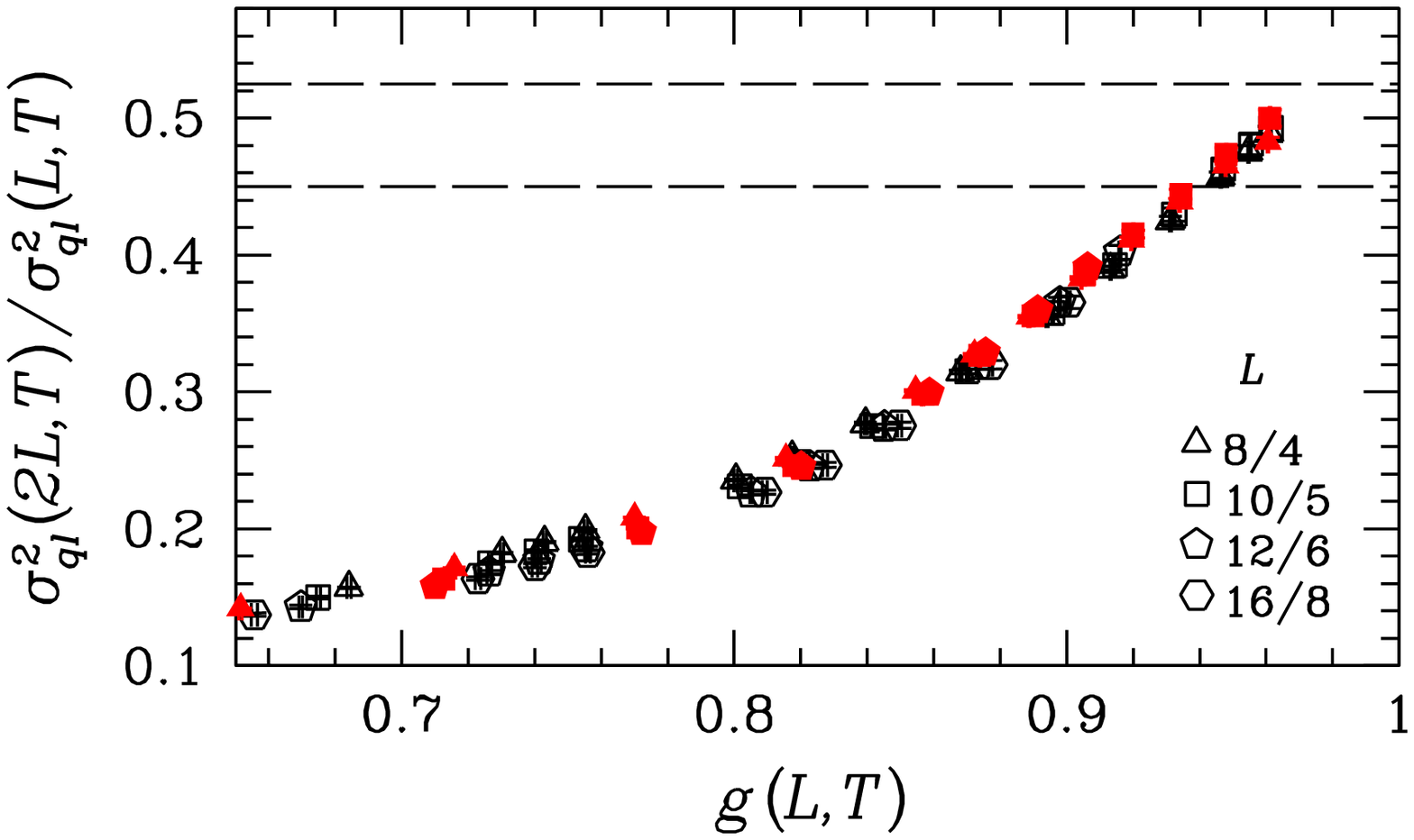}

\caption{(Color online)
Comparison of the finite-size scaling function of the link-overlap
variance $\sigma^2_{ql}$ for the hierarchical model (top) and the
3D EA model (bottom). For both MK an EA models the results for the
Gaussian (open/black symbols) and the irrational disorder (full/red
symbols) are shown. The broken line (the range between the broken
lines, respectively) correspond(s) to the expected value of the
scaling function for $g \to 1$ in the thermodynamic limit within the
droplet model.}
\label{fig:qlink_scaling}
\end{figure}

In Fig.~\ref{fig:qlink_scaling} we show results for the link-overlap
variance.  The top panel shows data for the MK model with Gaussian
and irrational disorder distributions. In contrast to the other FSS
functions discussed so far we find sizable scaling corrections. The
data are less conclusive but there is a clear trend that the curves
for the two different distributions converge to a single master
curve. The bottom panel shows data for the aforementioned disorder
distributions in the 3D EA model.  The data are similar to the MK
case and show small corrections to scaling.  The broken lines (or the
region between them) correspond(s) to the expected value of the
scaling function at $g = 1$ in the thermodynamic limit under droplet
scaling assumptions (here we use $\theta=0.20(5)$ and $d-d_s=0.42(3)$
\cite{palassini:00} in the 3D EA case; $\theta=0.27$ and $d_s= d -
1 = 2$ in the MK case).  Neither the droplet nor the RSB picture
(where $F_{\sigma^2_{q_l}} \to 1$) extrapolate to a consistent value
for $F_{\sigma^2_{q_l}}$ in the limit $g \to g(L,T=0)$,  in agreement
with Refs.~\cite{palassini:00,krzakala:00}.

\paragraph*{Summary and discussion.---}
\label{sec:conclusions}

Studying the behavior of several FSS functions we have found evidence
for a universal scaling behavior {\em in} the spin-glass phase. The
existence of these universal FSS functions allows us to perform for
the first time a precise comparison of results in the spin-glass phase
between different coupling distributions.  We find that neither the
simple droplet nor the RSB picture extrapolate to consistent values
for the scaling functions in the zero-temperature limit.  For both
pictures it can  be argued that these inconsistencies might be due
to scaling corrections \cite{middleton:01,marinari:00b}. Our
results suggest that in such a case not only the scaling behavior in
the thermodynamic limit is universal, but also the leading scaling
corrections. Although our results do not allow us to discriminate
between the different scenarios that have been proposed for the 
nature of the spin-glass phase, they allow for a new, parameter-free 
way of looking at the problem.

\begin{acknowledgments}

We thank F.~Krz\c{a}ka{\l}a, M.~A.~Moore and A.~P.~Young for 
discussions.  The simulations have been performed at ETH Z\"urich, 
LPTMS Orsay, ITP Bern and Roma I.  
H.G.K.~was supported by the SNF (Grant No.~PP002-114713) and 
T.J.~by EEC's FP6 IST contract under IST-034952.

\end{acknowledgments}

\vspace*{-0.6cm}

\bibliography{refs,comments}

\begin{thebibliography}{29}
\bibitem[{com({\natexlab{a}})}]{comment:bc}
\bibinfo{note}{The boundary conditions for the MK model involve just a single
  link and thus have a very small influence.}
\bibitem[{com({\natexlab{b}})}]{comment:fp}
\bibinfo{note}{In the (scaling) [RSB] picture there is (one fixed point) [a
  line of fixed points] with ($g^*=1)$ [$g_{\rm min}\leq g^*\leq 1$].}
\bibitem[{com({\natexlab{c}})}]{comment:cluster}
\bibinfo{note}{For the link-diluted model in 4D we use additional cluster moves
  \cite{joerg:06} to speed up equilibration.}
\end{thebibliography}

\end{document}